\documentclass[conference]{IEEEtran}
\IEEEoverridecommandlockouts

\usepackage{amsmath,amssymb,amsfonts}
\usepackage{algorithmic}
\usepackage{graphicx}
\usepackage{textcomp}
\usepackage{xcolor}
\usepackage{url}
\usepackage{orcidlink}
\usepackage{multirow}
\usepackage{subcaption}
\usepackage{ulem}
\usepackage{enumitem}
\usepackage[noadjust]{cite}

\def\BibTeX{{\rm B\kern-.05em{\sc i\kern-.025em b}\kern-.08em
    T\kern-.1667em\lower.7ex\hbox{E}\kern-.125emX}}
\begin{document}

\title{Mixture of Experts Approaches in Dense Retrieval Tasks}

\author{\IEEEauthorblockN{Effrosyni Sokli\orcidlink{0009-0003-5388-2385}}
\IEEEauthorblockA{\textit{University of Milano-Bicocca} \\
Milan, Italy \\
effrosyni.sokli@unimib.it}
\and
\IEEEauthorblockN{Pranav Kasela\orcidlink{0000-0003-0972-2424}}
\IEEEauthorblockA{\textit{University of Milano-Bicocca} \\
Milan, Italy \\
pranav.kasela@unimib.it}
\and
\IEEEauthorblockN{Georgios Peikos\orcidlink{0000-0002-2862-8209}}
\IEEEauthorblockA{\textit{University of Milano-Bicocca} \\
Milan, Italy \\
georgios.peikos@unimib.it}
\and
\IEEEauthorblockN{Gabriella Pasi\orcidlink{0000-0002-6080-8170}}
\IEEEauthorblockA{\textit{University of Milano-Bicocca} \\
Milan, Italy \\
gabriella.pasi@unimib.it}
}

\maketitle

\begin{abstract}
Dense Retrieval Models (DRMs) are a prominent development in Information Retrieval (IR).
A key challenge with these neural Transformer-based models is that they often struggle to generalize beyond the specific tasks and domains they were trained on.
To address this challenge, prior research in IR incorporated the Mixture-of-Experts (MoE) framework within each Transformer layer of a DRM, which, though effective, substantially increased the number of additional parameters.
In this paper, we propose a more efficient design, which introduces a single MoE block (\textsc{SB-MoE}) after the final Transformer layer.
To assess the retrieval effectiveness of \textsc{SB-MoE}, we perform an empirical evaluation across three IR tasks.
Our experiments involve two evaluation setups, aiming to assess both in-domain effectiveness and the model's zero-shot generalizability.
In the first setup, we fine-tune \textsc{SB-MoE} with four different underlying DRMs on seven IR benchmarks and evaluate them on their respective test sets.
In the second setup, we fine-tune \textsc{SB-MoE} on MSMARCO and perform zero-shot evaluation on thirteen BEIR datasets.
Additionally, we perform further experiments to analyze the model's dependency on its hyperparameters (i.e., the number of employed and activated experts) and investigate how this variation affects \textsc{SB-MoE}'s performance.
The obtained results show that \textsc{SB-MoE} is particularly effective for DRMs with lightweight base models, such as TinyBERT and BERT-Small, consistently exceeding standard model fine-tuning across benchmarks.
For DRMs with more parameters, such as BERT-Base and Contriever, our model requires a larger number of training samples to achieve improved retrieval performance.
Our code is available online at: \url{https://github.com/FaySokli/SB-MoE}.
\end{abstract}


\section{Introduction}
Neural Information Retrieval (IR) models, including Dense Retrieval Models (DRMs), have shown the potential to improve retrieval performance compared to sparse lexicon-based models such as BM25 \cite{mitra2018introduction}.
DRMs can be trained to capture the semantic context of queries and documents.
Nevertheless, their training requires large labeled datasets, often leading to a trade-off between generalizability and task-specific performance, as DRMs can struggle to robustly adapt to different tasks or domains without additional fine-tuning \cite{thakur2021beir}.

In this paper, we investigate the performance of an enhanced bi-encoder DRM architecture that leverages Mixture-of-Experts (MoE) \cite{jacobs1991adaptive}, and we compare it with the original DRM across various dense retrieval tasks. 
Since MoE consists of multiple sub-networks of experts, we assess whether their integration with DRMs allows each expert to capture meaningful textual characteristics (e.g., domain, text complexity, etc.), thereby enriching the final embedding representations of queries and documents.
Previous work in IR has integrated MoE within each Transformer layer of the DRM \cite{Dai2022MixtureOE,ma2023cot,DBLP:conf/iir/SokliRP24}, aiming to enhance the performance and generalizability of the proposed models in downstream tasks.
These MoE approaches yield performance improvements; however, they also significantly increase the number of additional parameters.
In this work, we add a single MoE block (\textsc{SB-MoE}) on the output embeddings of the final Transformer layer of the underlying DRM, reducing the number of additional parameters and maintaining efficiency.
\textsc{SB-MoE} consists of multiple pairs of Feed-Forward Networks (FFNs), where each pair acts as a unique expert.
The selection of the experts for each input is determined by a gating function, which is a neural network trained in an \textit{unsupervised} manner to automatically assign an importance weight to each expert.
The weights indicate how relevant each expert is to the given input, and along with a predefined pooling strategy, determine the model's final prediction.
Hence, \textsc{SB-MoE} optimizes each expert individually and dynamically selects and aggregates their outputs to tailor predictions to the input query or document representation.

We investigate the impact of integrating a single MoE block into DRMs, refining both the query and document embeddings.
We conduct an empirical evaluation on the three IR tasks of passage retrieval, open domain Q\&A, and domain-specific academic search.
To this end, we report results from two distinct evaluation setups designed to assess \textsc{SB-MoE}’s in-domain and task-specific effectiveness as well as its generalizability.
In the first setup, we fine-tune \textsc{SB-MoE} on the training set of seven different IR benchmarks, and evaluate it on the respective test set.
We compare its performance to that of the original DRM, trained in the same manner.
In the second setup, we fine-tune \textsc{SB-MoE} on the training set of MSMARCO and then conduct a zero-shot evaluation on the test sets of thirteen BEIR datasets to assess \textsc{SB-MoE}'s generalizability.
Retrieval effectiveness is compared against the fine-tuned version of the original DRM, following the same evaluation protocol.
\begin{figure*}[t]
    \centering
    \includegraphics[width=0.6\linewidth]{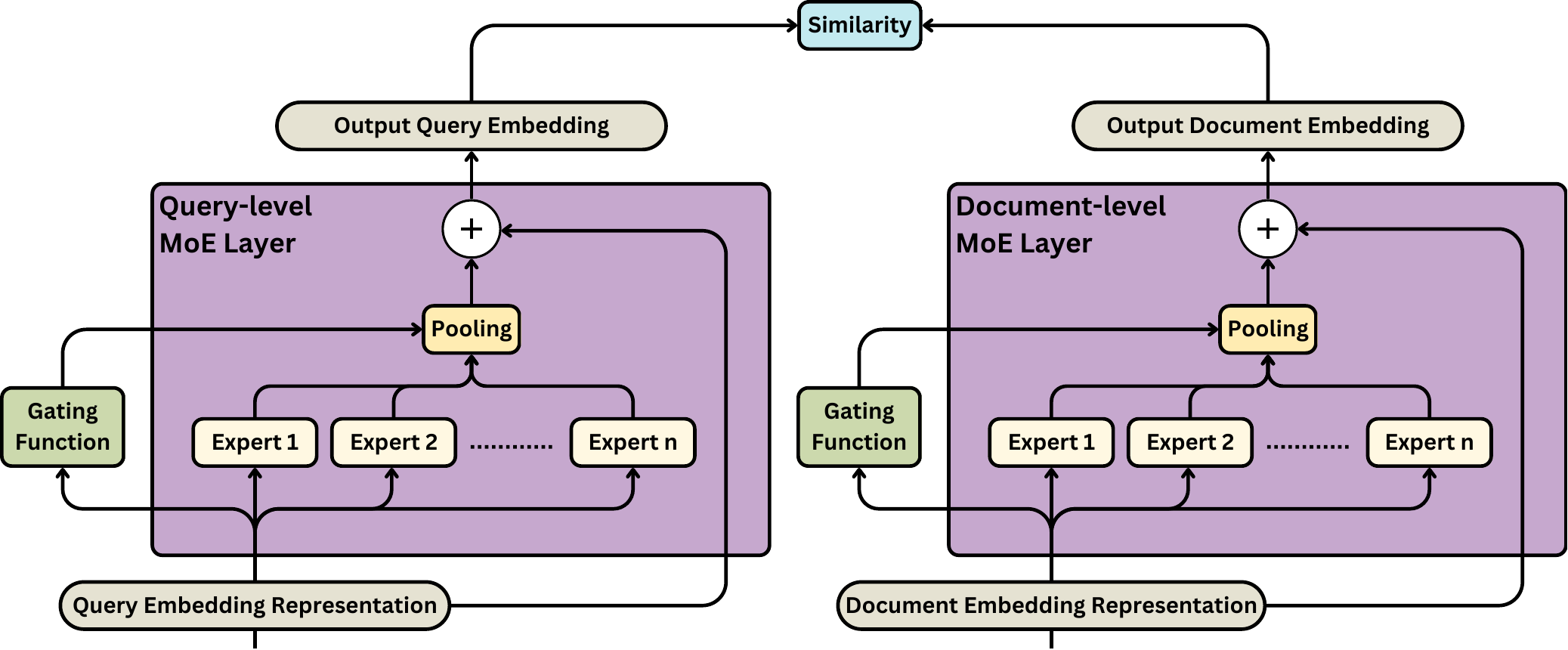}
    \caption{\textsc{SB-MoE} with its three main parts: (i) the experts, (ii) the gating function, and (iii) the pooling module.}
    \label{fig:moe-arch}
\end{figure*}
The contributions of this work are the following: 
\begin{enumerate}
    \item The introduction of a \textit{modular MoE framework} integrated into a dense retrieval architecture; the framework takes as input the query and the document embeddings outputted from the final Transformer layer of the underlying DRM;
    \item An \textit{experimental analysis} on four different DRMs to investigate our model's retrieval effectiveness and the impact of its hyperparameters (i.e., the number of employed and activated experts during training and inference).
\end{enumerate}

\section{Related Work}

DRMs often outperform IR lexicon-based models (e.g., BM25 \cite{DBLP:conf/trec/RobertsonWJHG94}), since they can capture the semantic context of queries and documents by projecting them in a shared dense vector space.
Similarity functions are used to estimate relevance and rank documents for a given query, based on the query and document embeddings \cite{gao-callan-2022-unsupervised,kamalloo2024resources,yu-etal-2022-coco}.
We experiment with four DRMs, each using a different base model to encode queries and documents, specifically leveraging Contriever \cite{DBLP:journals/tmlr/IzacardCHRBJG22}, BERT-Base \cite{devlin-etal-2019-bert}, BERT-Small \cite{turc2019well}, and TinyBERT \cite{jiao-etal-2020-tinybert}.
Contriever (110M) is a state-of-the-art BERT-based model that is pre-trained using \textit{contrastive learning}, a Machine Learning technique that uses pairs of positive and negative examples to learn meaningful and distinctive representations of queries and documents.
BERT-Base (110M), a foundational Transformer-based model, learns deep contextualized representations through masked language modeling and next-sentence prediction.
BERT-Small (29M) leverages \textit{knowledge distillation} \cite{romero2014fitnets} to transfer knowledge from its larger counterpart, BERT, to a more compact version, maintaining a similar architecture but with fewer parameters to balance efficiency and performance.
TinyBERT (14.5M) is another tinier and distilled version of BERT, reducing training times and computational expenses.
We leverage the above models with varying parameter counts and pre-training strategies to evaluate the impact of the integrated MoE block on DRM performance across retrieval tasks.

A common downside of DRMs is their continuous need for domain and task adaptation, which often leads to low generalizability without additional fine-tuning \cite{liu2024robust,sidiropoulos2022analysing}.
A potential way to improve a model's generalizability is MoE \cite{jacobs1991adaptive}, with several approaches proposed in the literature \cite{DBLP:conf/iclr/LiSYWRCZ023,guo2024came,shen2024mixtureofexperts,DBLP:journals/corr/abs-2412-11864}.
The MoE framework can handle multiple types of data and tasks \cite{collobert2001parallel,li-etal-2021-multi-task-dense}, and it has been implemented in different scenarios, such as classification tasks \cite{eigen2013learning}, and multilingual machine translation \cite{shazeer2017outrageously}.
In IR, MoE has been used for tasks such as passage retrieval \cite{ma2023cot,guo2024came}, and Q\&A \cite{Dai2022MixtureOE,kasela2024desire-me}; the approaches proposed in \cite{Dai2022MixtureOE}, \cite{ma2023cot}, and \cite{guo2024came} replace the feed-forward block of the underlying Transformer-based model's layers with multiple FFNs, each acting as a distinct expert.
While these approaches benefit the underlying model, they often substantially increase the overall number of parameters.
In \cite{kasela2024desire-me}, the authors extend the underlying DRM by applying a MoE block solely to the query embedding representation, thereby only partially exploring the potential of integrating MoE into the DRM architecture.

To address the limitations of existing studies and to further explore the potential of MoE implementations in IR, we: 
\begin{enumerate}
    \item Expand the DRM architecture by adding a \textbf{single} MoE block (\textsc{SB-MoE}) instead of integrating multiple MoE blocks within every Transformer layer;
    \item Apply the MoE block on \textbf{both} query and document representations produced by the underlying DRM; and
    \item Evaluate the retrieval effectiveness of \textsc{SB-MoE} on three IR tasks, i.e., passage retrieval, open-domain Q\&A, and domain-specific academic search.
\end{enumerate}

\section{Methodology}\label{sec:method}

\textsc{SB-MoE} builds upon a bi-encoder DRM architecture \cite{reimers-gurevych-2019-sentence}, aiming to maintain efficiency while improving the computation of relevance scores through a similarity function (e.g., dot product).
Bi-encoders typically allow for the independent encoding of the query and document embeddings, leading to efficient retrieval via precomputed embeddings.
While it is possible to employ separate encoders \cite{karpukhin2020dense}, using the same base model for both queries and documents improves the DRM's robustness, without significantly affecting performance \cite{DBLP:journals/tmlr/IzacardCHRBJG22,reimers-gurevych-2019-sentence}. 
Figure \ref{fig:moe-arch} illustrates the proposed model's architecture.
\textsc{SB-MoE} consists of: 
(1) a \textit{DRM} (not shown in Figure \ref{fig:moe-arch}), which generates the query and document embedding representations (depicted as ``Query Embedding Representation" and ``Document Embedding Representation" in Figure \ref{fig:moe-arch}).
(2) \textit{the experts}, which are added after the output of the final Transformer layer of the underlying DRM and act on both query and document embeddings; 
(3) \textit{the gating function}, which is trained in an unsupervised manner to select the best expert for each given input; and 
(4) \textit{the pooling module} used in the final stage to aggregate the experts' representations and produce the final embedding to be used for similarity estimation between queries and documents.

\textit{The experts} receive the input query or document embedding directly from the underlying DRM (Figure \ref{fig:moe-arch}).
The output is constituted by $n$ distinct and enhanced representations for each input query or document embedding, where $n$ is the total number of experts.
The outputs from all experts are subsequently aggregated into a single representation for each input query or document embedding, guided by the pooling module.
\textit{The gating function} takes as input the query or document embedding as outputted from the underlying DRM.
Subsequently, it produces an $n$-dimensional vector that expresses the likelihood of each expert being suited to handle the given input query or document embedding.
These weights indicate the importance of each expert’s contribution to the input query or document embedding.
Finally, \textit{the pooling module} aggregates all experts' outputs to form the final embedding representation to be used for similarity estimation between queries and documents.
During training, we rely on noisy Top-1 gating, as proposed by Shazeer et al. \cite{shazeer2017outrageously}, to ensure that our model explores every expert.
During inference, our model can operate with two different pooling strategies, resulting in two distinct variants, namely \textsc{SB-MoE}$_{\text{TOP-1}}$ and \textsc{SB-MoE}$_{\text{ALL}}$.
These variants represent different approaches to expert aggregation, used to explore the impact of expert contributions on retrieval effectiveness.
Hereafter, we use the model name ``\textsc{SB-MoE}" to refer to both variants. 

In detail, \textsc{SB-MoE}$_{\text{TOP-1}}$ relies on Top-1 gating \cite{DBLP:conf/nips/ZhouLLDHZDCLL22}, which selects only the output of the expert to which the gating function assigned the highest score.
Formally, let $x$ be the input embedding and $f_{i}(x)$ the output of function $f_{i}$ learned by the $i$-th expert.
If $g_{i}(x)$ is the weight assigned to the $i$-th expert by the gating function given the input query or document embedding $x$, then:
\begin{equation} \label{eq:gate}
    m = \arg \max_{i=1,...,n}(g_i(x)),
\end{equation}
\begin{equation} \label{eq:expert_1}
    y = f_m(x), 
\end{equation}
where $n$ is the total number of experts, and $m$ represents the most appropriate expert as indicated by the gating function.
This leads to the final refined representation $y$ of the given query or document embedding.

The second variant (\textsc{SB-MoE}$_{\text{ALL}}$) calculates probability scores from the gating function's output vector through a softmax normalization \cite{JordanHierarchical1994}, and based on them it produces the final output embedding.
Formally, the final refined query or document representation $y$ is the weighted sum of the outputted embeddings from all $n$ employed experts:
\begin{equation} \label{eq:experts_all}
    y = \sum_{i=1}^{n} f_i(x) \cdot g_i(x)
\end{equation}
This pooling strategy leverages all employed experts and utilizes their distinct specializations collectively to refine the query and document representations for similarity estimation.

\section{Experimental Setup}\label{sec:experiments}
\begin{table*}[t]
\caption{Results on all datasets. Metrics refer to Recall@100 and nDCG@10. Symbol * indicates a statistically significant difference over \textit{fine-tuned}, calculated using the ASPIRE toolkit~\cite{DBLP:journals/corr/abs-2412-15759}. Best results for each dataset are in \textbf{bold}.}
\label{tab:results_v2}
\centering
\resizebox{1\linewidth}{!}{%
\begin{tabular}{ll|rrrr|rrrr|rrrr|rrrrrrrr}
\multicolumn{2}{r}{}        & \multicolumn{4}{c}{TinyBERT} & \multicolumn{4}{c}{BERT-Small} & \multicolumn{4}{c}{BERT-Base} & \multicolumn{4}{c}{Contriever} \\
\multicolumn{2}{r|}{} & \rotatebox{60}{\textsc{fine-tuned}} & \rotatebox{60}{\textsc{random-gate}} & \rotatebox{60}{\textsc{SB-MoE}$_{\text{TOP-1}}$} & \rotatebox{60}{\textsc{SB-MoE}$_{\text{ALL}}$} & \rotatebox{60}{\textsc{fine-tuned}} & \rotatebox{60}{\textsc{random-gate}} & \rotatebox{60}{\textsc{SB-MoE}$_{\text{TOP-1}}$} & \rotatebox{60}{\textsc{SB-MoE}$_{\text{ALL}}$} & \rotatebox{60}{\textsc{fine-tuned}} & \rotatebox{60}{\textsc{random-gate}} & \rotatebox{60}{\textsc{SB-MoE}$_{\text{TOP-1}}$} & \rotatebox{60}{\textsc{SB-MoE}$_{\text{ALL}}$} & \rotatebox{60}{\textsc{fine-tuned}} & \rotatebox{60}{\textsc{random-gate}} & \rotatebox{60}{\textsc{SB-MoE}$_{\text{TOP-1}}$} & \rotatebox{60}{\textsc{SB-MoE}$_{\text{ALL}}$} \\ \hline
\multirow{2}{*}{MSMARCO}    & \textit{recall} & .682 & .686*           & \textbf{.688*}           & \textbf{.688*}           & .734 & .732\phantom{0} & \textbf{.736*} & \textbf{.736*} & .790 & .788\phantom{0} & .790\phantom{0} & \textbf{.791}\phantom{0} & \textbf{.850} & .838* & .839* & .839* \\
                            & \textit{nDCG}   & .244 & .243*           & .245*                    & \textbf{.246*}           & .254 & .253*           & \textbf{.255*} & \textbf{.255*} & \textbf{.293} & .286* & .289* & .290* & \textbf{.321} & .308* & .309* & .309* \\ \hline
\multirow{2}{*}{TREC DL 19} & \textit{recall} & .421 & .408\phantom{0} & .421\phantom{0}          & \textbf{.422}\phantom{0} & .464 & .455\phantom{0} & .464\phantom{0} & \textbf{.466}\phantom{0} & .530 & .509* & .528\phantom{0} & \textbf{.531}\phantom{0} & \textbf{.605} & .583* & .582* & .584* \\
                            & \textit{nDCG}   & .439 & .434\phantom{0} & \textbf{.440}\phantom{0} & .438\phantom{0}          & .430 & .436\phantom{0} & .437\phantom{0} & \textbf{.440}\phantom{0} & .506 & .503\phantom{0} & .502\phantom{0} & \textbf{.507}\phantom{0} & .540 & .533\phantom{0} & \textbf{.545}\phantom{0} & .539\phantom{0} \\ \hline
\multirow{2}{*}{TREC DL 20} & \textit{recall} & .521 & .516\phantom{0} & \textbf{.524}\phantom{0} & .516\phantom{0}          & .573 & .565\phantom{0} & \textbf{.574}\phantom{0} & .570\phantom{0} & .614 & .615\phantom{0} & .617\phantom{0} & \textbf{.620}\phantom{0} & .678 & .665\phantom{0} & \textbf{.683}\phantom{0} & \textbf{.683}\phantom{0} \\
                            & \textit{nDCG}   & .466 & .466\phantom{0} & .452\phantom{0}          & \textbf{.468}\phantom{0} & .449 & .450\phantom{0} & \textbf{.456}\phantom{0} & .453\phantom{0} & \textbf{.516} & .508\phantom{0} & .513\phantom{0} & \textbf{.516}\phantom{0} & \textbf{.543} & .504* & \textbf{.543}\phantom{0} & .535\phantom{0} \\ \hline
\multirow{2}{*}{NQ}         & \textit{recall} & .722 & .721\phantom{0} & .723\phantom{0}          & \textbf{.725*}           & .760 & .764* & \textbf{.774*} & .763* & .881 & .879\phantom{0} & .880\phantom{0} & \textbf{.883}\phantom{0} & \textbf{.934} & .926* & .930* & .932\phantom{0} \\
                            & \textit{nDCG}   & .261 & .264*           & .263*                    & \textbf{.271*}           & .215 & .219* & \textbf{.227*} & .218* & .350 & .351\phantom{0} & .352* & \textbf{.356*} & \textbf{.426} & .403* & .416* & .416* \\ \hline
\multirow{2}{*}{HotpotQA}   & \textit{recall} & .465 & .458*           & .464\phantom{0}          & \textbf{.477*}           & .587 & .583* & .588\phantom{0} & \textbf{.595*} & \textbf{.699} & .695* & .691* & .695* & \textbf{.862} & .855* & .853* & .861\phantom{0} \\
                            & \textit{nDCG}   & .243 & .235*           & .236*                    & \textbf{.252*}           & .310 & .309\phantom{0} & .310\phantom{0} & \textbf{.317*} & \textbf{.444} & .430* & .425* & .431* & \textbf{.672} & .653* & .653* & .667* \\ \hline
\multirow{2}{*}{PS}         & \textit{recall} & .282 & .284*           & .284*                    & \textbf{.291*}           & .335 & .335\phantom{0} & .334\phantom{0} & \textbf{.337*} & \textbf{.396} & .395\phantom{0} & .394* & \textbf{.396}\phantom{0} & \textbf{.483} & .471* & .479* & \textbf{.483}\phantom{0} \\
                            & \textit{nDCG}   & .145 & .146\phantom{0} & .145\phantom{0}          & \textbf{.149*}           & .164 & .164\phantom{0} & .163\phantom{0} & \textbf{.165}\phantom{0} & \textbf{.201} & .200\phantom{0} & .200\phantom{0} & \textbf{.201}\phantom{0} & \textbf{.251} & .243* & .250\phantom{0} & \textbf{.251}\phantom{0} \\ \hline
\multirow{2}{*}{CS}         & \textit{recall} & .327 & .328\phantom{0} & .328\phantom{0}          & \textbf{.332*}           & .324 & .326\phantom{0} & .328* & \textbf{.330*} & \textbf{.387} & .383* & .382* & .383* & .437 & .435\phantom{0} & .435\phantom{0} & \textbf{.438}\phantom{0} \\
                            & \textit{nDCG}   & .172 & .171\phantom{0} & .172\phantom{0}          & \textbf{.175*}           & .166 & .166\phantom{0} & .168* & \textbf{.169*} & .197 & .196\phantom{0} & .197\phantom{0} & \textbf{.198*} & \textbf{.224} & \textbf{.224}\phantom{0} & .222* & .223\phantom{0} \\
\end{tabular}
}
\end{table*}

This section presents the experimental setup of the empirical evaluation conducted to answer the following research questions (RQs):

\begin{enumerate}[noitemsep, topsep=0ex, align=left,label={\bfseries RQ\arabic*:}]
    \item How does \textsc{SB-MoE} compare, in terms of retrieval effectiveness, to the fine-tuned version of the underlying DRM?
    \item To what extent can the observed performance gains be attributed to the learned gating function and to the experts' ability to capture useful textual characteristics?
    \item Can \textsc{SB-MoE} generalize well across multiple domains and tasks without additional fine-tuning?
\end{enumerate}
In answering these research questions, we compare our approach to several baselines that are analyzed in Section \ref{sec:baselines}.

\subsection{Baselines and Metrics} \label{sec:baselines}
We integrate the single MoE block within four DRMs with different base models\footnote{Available on HuggingFace: 
\href{https://huggingface.co/facebook/contriever}{Contriever}, \href{https://huggingface.co/google-bert/bert-base-uncased}{BERT-Base}, \href{https://huggingface.co/google/bert_uncased_L-4_H-512_A-8}{BERT-Small} and \href{https://huggingface.co/huawei-noah/TinyBERT_General_4L_312D}{TinyBERT}}.
We compare \textsc{SB-MoE}'s retrieval effectiveness to that achieved by the underlying DRM, fine-tuned on the same training data and hyperparameters.
We refer to these baseline experiments as \textsc{fine-tuned}, which serve to address RQ1.
Additionally, we employ another baseline, \textsc{random-gate}, in which random weights are assigned to the experts instead of relying on the learned gating function.
This baseline is designed to assess the contribution of the gating function and the experts' ability to capture useful textual characteristics for embedding refinement, thus addressing RQ2.
Finally, to assess \textsc{SB-MoE}'s generalizability (RQ3), we fine-tune all four DRMs on MSMARCO (a widely adopted benchmark for zero-shot retrieval evaluation \cite{kamalloo2024resources}), with and without the addition of MoE.
We then proceed with its zero-shot evaluation on the test sets of 13 BEIR \cite{thakur2021beir} datasets.
We refer to this baseline as \textsc{ft}$_{\text{MSMARCO}}$.
We evaluate the experimental results using nDCG@10 and Recall@100 (two metrics commonly used in the selected evaluation collections) for comparability.
Statistical significance has been evaluated based on two-sided paired Student's $t$-tests with Bonferroni multiple testing correction at significance levels of 0.05.

\subsection{Training Hyperparameters} 
To address our RQs, we employ 6 distinct experts for all models and datasets.
We also conduct further experiments where we vary the number of experts in the range of 3-12 with a step of 3.
Our selection of the number of employed experts is based on previous studies investigating the impact of expert count on model performance \cite{DBLP:conf/coling/LiHWYXJLLZ24,DBLP:conf/iclr/ZadouriUAELH24}, which reveal that a high number of experts does not necessarily lead to performance gains \cite{DBLP:conf/iclr/ChenZJLW23}, and set the number of experts in the range of 2-8 \cite{Dai2022MixtureOE,ma2023cot,wang-etal-2022-adamix}.
Following the architecture proposed by Houlsby et al. \cite{HoulsbyParameter2019}, each expert within the proposed MoE block consists of a down-projection FFN layer reducing the input dimension by half, followed by an up-projection FFN output layer that restores the vector dimension to match that of the input's embedding.
Their design also includes a skip connection, which we use to enable residual learning.
The gating function consists of a single hidden layer, which reduces the vector dimension by half, and an output layer, which is an FFN with the same size as the number of experts. 
We set the training batch size to 64 and the learning rate to $10^{-6}$ and $10^{-4}$ for the underlying DRM and the experts, respectively. 
We train TinyBERT for 30 epochs across all datasets, as its smaller size allows for faster training times. 
BERT-Small, BERT-Base, and Contriever are trained for 20 epochs due to resource constraints and longer training times, on all datasets except CS, where they are trained for 10 epochs since the collection's training queries are $\sim $3.5\% more than the second largest collection used.
We use 5\% of the training sets for validation and keep only the checkpoint with the lowest validation loss.
We set the random seed to 42 and use contrastive loss \cite{DBLP:journals/tmlr/IzacardCHRBJG22} with a temperature of 1 for all models except Contriever, where the authors report an optimal temperature of 0.05.

\subsection{Datasets}
In our experiments, we use seven publicly available datasets: 
    \textit{MSMARCO v1 passage corpus} \cite{DBLP:conf/nips/NguyenRSGTMD16}, which consists of approximately 8.8 million passages, with 532k training queries with 1.1 average relevance judgments per query and 7k test queries;
    \textit{TREC Deep Learning Tracks of 2019 and 2020 (TREC DL 19 \& 20)} \cite{DBLP:journals/corr/abs-2003-07820,DBLP:journals/corr/abs-2102-07662}, which share the same corpus with MSMARCO but differ in terms of the average relevance judgments per query (215 and 211, respectively) and number of test queries (43 and 54, respectively);
    \textit{Natural Questions (NQ)} \cite{kwiatkowski2019NQ}, an open-domain Q\&A dataset from the BEIR benchmark \cite{thakur2021beir} based on a corpus consisting of 2.6 million Wikipedia passages, with 132k training queries with 1.2 average relevance judgments per query and 3.5k test queries; 
    \textit{HotpotQA} \cite{yang-etal-2018-hotpotqa}, another Q\&A dataset from the BEIR benchmark with multihop questions that also relies on a Wikipedia-based corpus of 5.2 million documents, with 85k training queries with 2 average relevance judgments per query and 7.4k test queries;
    Two datasets from the Multi-Domain Benchmark proposed by Bassani et al. \cite{bassani2022multidomain}, namely \textit{Political Science (PS)} and \textit{Computer Science (CS)}, for the task of domain-specific academic search.
    Both collections consist of 4.8M documents, with PS containing 160k training queries with an average of 3.8 relevance judgments per query and 5.7k testing queries, while CS contains 550k training queries with an average of 3.25 relevance judgments per query and 6.5k testing queries.
We use these datasets with distinct characteristics to evaluate \textsc{SB-MoE} under different conditions and to study its consistency across various training sample sizes and retrieval tasks.
Additionally, we fine-tune \textsc{SB-MoE} once on MSMARCO and then conduct a \textit{zero-shot} evaluation on the test sets of 13 datasets of the BEIR collection \cite{thakur2021beir}.

\section{Results} \label{sec:results}
\begin{table*}[t]
\caption{Results on the \textit{zero-shot} evaluation of \textsc{SB-MoE} on the test sets of 13 BEIR datasets. Comparison of our model fine-tuned on MSMARCO with all four underlying DRMs. Metrics refer to Recall@100 and nDCG@10. Symbol * indicates a statistically significant difference over \textsc{ft}$_{\text{MSMARCO}}$. Best results for each model are indicated with \textbf{bold} text.}
\label{tab:results_zero}
\centering
\resizebox{1\linewidth}{!}{%
\begin{tabular}{ll|ll|ll|ll|ll|ll|ll|ll|ll|ll|ll|ll|ll|ll}
\multicolumn{2}{r|}{}       & \multicolumn{2}{c|}{covid} & \multicolumn{2}{c|}{nfcorpus} & \multicolumn{2}{c|}{NQ} & \multicolumn{2}{c|}{HotpotQA} & \multicolumn{2}{c|}{fiqa} & \multicolumn{2}{c|}{arguana} & \multicolumn{2}{c|}{touche} & \multicolumn{2}{c|}{quora} & \multicolumn{2}{c|}{dbpedia} & \multicolumn{2}{c|}{scidocs} & \multicolumn{2}{c|}{fever} & \multicolumn{2}{c|}{cfever} & \multicolumn{2}{c}{scifact}  \\
Retriever & Variant & \multicolumn{1}{l}{\rotatebox{60}{\textit{recall}}} & \multicolumn{1}{l|}{\rotatebox{60}{\textit{nDCG}}} & \multicolumn{1}{l}{\rotatebox{60}{\textit{recall}}} & \multicolumn{1}{l|}{\rotatebox{60}{\textit{nDCG}}} & \multicolumn{1}{l}{\rotatebox{60}{\textit{recall}}} & \multicolumn{1}{l|}{\rotatebox{60}{\textit{nDCG}}} & \multicolumn{1}{l}{\rotatebox{60}{\textit{recall}}} & \multicolumn{1}{l|}{\rotatebox{60}{\textit{nDCG}}} & \multicolumn{1}{l}{\rotatebox{60}{\textit{recall}}} & \multicolumn{1}{l|}{\rotatebox{60}{\textit{nDCG}}} & \multicolumn{1}{l}{\rotatebox{60}{\textit{recall}}} & \multicolumn{1}{l|}{\rotatebox{60}{\textit{nDCG}}} & \multicolumn{1}{l}{\rotatebox{60}{\textit{recall}}} & \multicolumn{1}{l|}{\rotatebox{60}{\textit{nDCG}}} & \multicolumn{1}{l}{\rotatebox{60}{\textit{recall}}} & \multicolumn{1}{l|}{\rotatebox{60}{\textit{nDCG}}} & \multicolumn{1}{l}{\rotatebox{60}{\textit{recall}}} & \multicolumn{1}{l|}{\rotatebox{60}{\textit{nDCG}}} & \multicolumn{1}{l}{\rotatebox{60}{\textit{recall}}} & \multicolumn{1}{l|}{\rotatebox{60}{\textit{nDCG}}} & \multicolumn{1}{l}{\rotatebox{60}{\textit{recall}}} & \multicolumn{1}{l|}{\rotatebox{60}{\textit{nDCG}}} & \multicolumn{1}{l}{\rotatebox{60}{\textit{recall}}} & \multicolumn{1}{l|}{\rotatebox{60}{\textit{nDCG}}} & \multicolumn{1}{l}{\rotatebox{60}{\textit{recall}}} & \multicolumn{1}{l}{\rotatebox{60}{\textit{nDCG}}} \\ \hline
\multirow{3}{*}{TinyBERT}   & \textsc{ft}$_{\text{MSMARCO}}$    
& .058 & .338 & .200 & \textbf{.212} & .653 & .245 & .467 & .320 & .368 & .129 & .836 & .196 & .321 & .131 & .968 &       .785 & .288 & .223 & .235 & \textbf{.102} & \textbf{.780} & .447 & \textbf{.386} & \textbf{.149} & .706 & .365  \\
 & \textsc{SB-MoE}$_{\text{TOP-1}}$ 
& \textbf{.061} & .348 & .199 & .209 & \textbf{.658*} & \textbf{.250*} & .469 & .320 & .377* & .131 & .837 & .201* & .329 & .142* & \textbf{.969} & \textbf{.786} & \textbf{.289} & .220 & .237 & .101 & .775* & .442* & .368* & .140* & \textbf{.725*} & .377*  \\
 & \textsc{SB-MoE}$_{\text{ALL}}$  
& \textbf{.061} & \textbf{.356} & \textbf{.204} & \textbf{.212} & .657* & \textbf{.250*} & \textbf{.470*} & \textbf{.326*} & \textbf{.379*} & \textbf{.133*} & \textbf{.839} & \textbf{.202*} & \textbf{.337} & \textbf{.143*} & \textbf{.969} & \textbf{.786} & .286 & \textbf{.224} & \textbf{.238} & .101 & .777 & \textbf{.452*} & .368* & .138* & \textbf{.725*} & \textbf{.378*}  \\ \hline
\multirow{3}{*}{BERT-Small} & \textsc{ft}$_{\text{MSMARCO}}$    
& .038 & .305 & .213 & .228 & .706 & .210 & \textbf{.607} & \textbf{.445} & .370 & .141 & .967 & \textbf{.371} & .230 & .049 & .976   & .794 & .340 & .221 & \textbf{.282} & \textbf{.119} & \textbf{.698} & \textbf{.314} & .318 & .100 & .776 & .424  \\
 & \textsc{SB-MoE}$_{\text{TOP-1}}$ 
& \textbf{.040} & .303 & .213 & .227 & \textbf{.707} & \textbf{.213*} & .602* & .437* & .386* & .146* & \textbf{.968} & .363* & \textbf{.240} & \textbf{.050} & .976   & .788* & \textbf{.342} & \textbf{.222} & .276* & .116* & .691* & .312* & \textbf{.322*} & \textbf{.104*} & .770 & .421  \\
 & \textsc{SB-MoE}$_{\text{ALL}}$   
& .039 & \textbf{.306} & \textbf{.216} & \textbf{.229} & .706 & .210 & .604 & .440* & \textbf{.387*} & \textbf{.147*} & .967 & .363* & .238 & .047 & \textbf{.977} & \textbf{.795} & \textbf{.342} & .219 & .278* & .116* & .692* & .311* & \textbf{.322*} & \textbf{.104*} & \textbf{.786} & \textbf{.432}  \\ \hline
\multirow{3}{*}{BERT-Base}  & \textsc{ft}$_{\text{MSMARCO}}$    
& \textbf{.079} & \textbf{.489} & \textbf{.238} & \textbf{.270} & \textbf{.829} & \textbf{.330} & \textbf{.633} & \textbf{.463} & \textbf{.544} & \textbf{.221} & .947 & .273 & \textbf{.398} & \textbf{.146} & .984 & .824 & \textbf{.407} & \textbf{.290} & .308 & .127 & \textbf{.911} & \textbf{.605} & \textbf{.489} & \textbf{.191} & \textbf{.863} & .540  \\
 & \textsc{SB-MoE}$_{\text{TOP-1}}$ 
& .077 & .470 & \textbf{.238} & .265 & .820* & .322* & .626* & .455* & .535* & .218 & \textbf{.954} & .281* & .378 & .143 & .984 & \textbf{.825} & .400 & .279* & \textbf{.309} & \textbf{.128} & .902* & .590* & .486 & .188* & .833* & .542  \\
 & \textsc{SB-MoE}$_{\text{ALL}}$   
& .078 & .474 & .237 & .266 & .820* & .323* & .627* & .457* & .539  & .218 & \textbf{.954} & \textbf{.283*} & .368* & .144 & \textbf{.985} & \textbf{.825} & .399* & .277* & .308 & .127 & .901* & .588* & .484* & .186* & .836* & \textbf{.543}  \\ \hline
\multirow{3}{*}{Contriever} & \textsc{ft}$_{\text{MSMARCO}}$    
& \textbf{.052} & \textbf{.427} & \textbf{.301} & \textbf{.333} & \textbf{.896} & \textbf{.376} & \textbf{.799} & \textbf{.650} & \textbf{.639} & \textbf{.291} & .920 & .302 & .346 & \textbf{.138} & \textbf{.993} & \textbf{.864} & \textbf{.547} & \textbf{.390} & \textbf{.388} & .166 & \textbf{.936} & \textbf{.681} & .485 & .185 & \textbf{.959} & \textbf{.679}  \\
 & \textsc{SB-MoE}$_{\text{TOP-1}}$ 
& .024* & .266* & .298  & .324* & .890* & .358* & .789* & .634* & .628* & .272* & \textbf{.977*} & .348* & .344          & .107* & .989 & .818* & .530* & .376* & .384 & .166          & \textbf{.936} & .614* & .505* & .197* & .942 & .665  \\
 & \textsc{SB-MoE}$_{\text{ALL}}$   
& .024* & .265* & .292* & .326* & .889* & .363* & .792* & .638* & .628* & .273* & .975* & \textbf{.350*} & \textbf{.347} & .116* & .989 & .818* & .529* & .377* & .384 & \textbf{.167} & .935          & .619* & \textbf{.508*} & \textbf{.198*} & .944 & .668  
\end{tabular}
}
\end{table*}
\textbf{RQ1.}
As shown in Table \ref{tab:results_v2}, \textsc{SB-MoE} (columns \textsc{SB-MoE}$_{\text{TOP-1}}$ and \textsc{SB-MoE}$_{\text{ALL}}$) exhibited a noticeable improvement in terms of nDCG@10 and Recall@100 compared to standard model fine-tuning (column \textsc{fine-tuned}), particularly for DRMs that have fewer parameters. 
This trend is evident across all datasets, showcasing that \textsc{SB-MoE} effectively enhances retrieval performance for DRMs with lightweight base models (i.e., TinyBERT and BERT-Small).
For instance, on TinyBERT and BERT-Small, our model leads to consistent performance gains in both metrics across all datasets, with a marked increase for BERT-Small in NQ, where \textsc{SB-MoE}$_{\text{TOP-1}}$ surpasses the \textsc{fine-tuned} version by 5.58\% in nDCG@10.
Despite its effectiveness in lightweight models, \textsc{SB-MoE} has a less pronounced impact when integrated into larger DRMs, such as BERT-Base and Contriever.
For example, on the MSMARCO and TREC DL 19 \& 20 datasets, the application of a single MoE block to BERT-Base resulted in retrieval effectiveness that was mostly on par with that of the fully \textsc{fine-tuned} model.
These results suggest that in models already equipped with a high number of parameters, the advantages of our model may not be so prominent, potentially due to redundancy when additional experts are employed.
Therefore, the integration of a single MoE block to DRMs is particularly advantageous for models with a smaller parameter size, where additional expert layers contribute to refining representations and improving robustness.

\textbf{RQ2.}
A key finding of our study is the importance of the gating function in effective expert selection.
To better assess the necessity of learned gating mechanisms, we compare \textsc{SB-MoE} against the \textsc{random-gate} baseline (column \textsc{random-gate} - Table \ref{tab:results_v2}), in which expert weights are assigned randomly.
We observe that \textsc{random-gate} performs comparably to the \textsc{fine-tuned} baseline (column \textsc{fine-tuned} - Table \ref{tab:results_v2}), indicating that simply adding experts, without any specific training strategy, is insufficient and does not enrich the generated query and document embeddings.
In contrast, \textsc{SB-MoE} (columns \textsc{SB-MoE}$_{\text{TOP-1}}$ and \textsc{SB-MoE}$_{\text{ALL}}$ - Table \ref{tab:results_v2}) consistently outperforms \textsc{random-gate}, highlighting the critical role of the gating function in dynamically selecting relevant experts.
Therefore, the observed performance improvements can be attributed to the learned gating function and the experts' ability to capture useful textual characteristics.

\textbf{RQ3.}
As presented in Table \ref{tab:results_zero}, \textsc{SB-MoE} (rows \textsc{SB-MoE}$_{\text{TOP-1}}$ and \textsc{SB-MoE}$_{\text{ALL}}$) exhibits consistent performance improvements in retrieval effectiveness across both DRMs with lightweight base models (i.e., TinyBERT and BERT-Small).
In detail, our model integrated with TinyBERT and fine-tuned on MSMARCO surpasses the \textsc{ft}$_{\text{MSMARCO}}$ baseline (row \textsc{ft}$_{\text{MSMARCO}}$) in 10 out of 13 BEIR datasets.
The performance gains of \textsc{SB-MoE}$_{\text{TOP-1}}$ and \textsc{SB-MoE}$_{\text{ALL}}$ on TinyBERT range from 0.10\% to 5.17\% in Recall@100 and from 0.13\% to 9.16\% in nDCG@10.
BERT-Small exhibits a performance increase in 9 datasets with gains ranging from 0.10\% to 5.26\% in Recall@100 and from 0.13\% to 4.26\% in nDCG@10.
These outcomes further support the findings presented in RQ1, highlighting the benefits of \textsc{SB-MoE} for lightweight DRMs and additionally showcasing its \textit{generalizability} across different domains and tasks.
For DRMs with larger parameter counts, our model performs on par with or outperforms \textsc{ft}$_{\text{MSMARCO}}$ in 6 datasets when integrated with BERT-Base and 5 datasets when integrated with Contriever.
The impact of the employed MoE block is less pronounced for these two DRMs; however, in certain cases, we observe performance improvements up to 6.2\% in Recall@100 and 15.89\% in nDCG@10 (e.g., Contriever on ArguAna), suggesting that \textsc{SB-MoE} can still provide substantial benefits even in DRMs with already strong baseline performance.
Overall, our observations suggest that \textsc{SB-MoE}, trained once and fine-tuned on MSMARCO, can generalize well across domains and tasks while maintaining low computational overhead, making it well-suited for low-resource environments with limited computational capacity.

\section{Hyperparameter Analysis}
\begin{figure}[t]
    \centering
    \begin{subfigure}[b]{0.47\textwidth}
        \includegraphics[width=\textwidth, height=5cm]{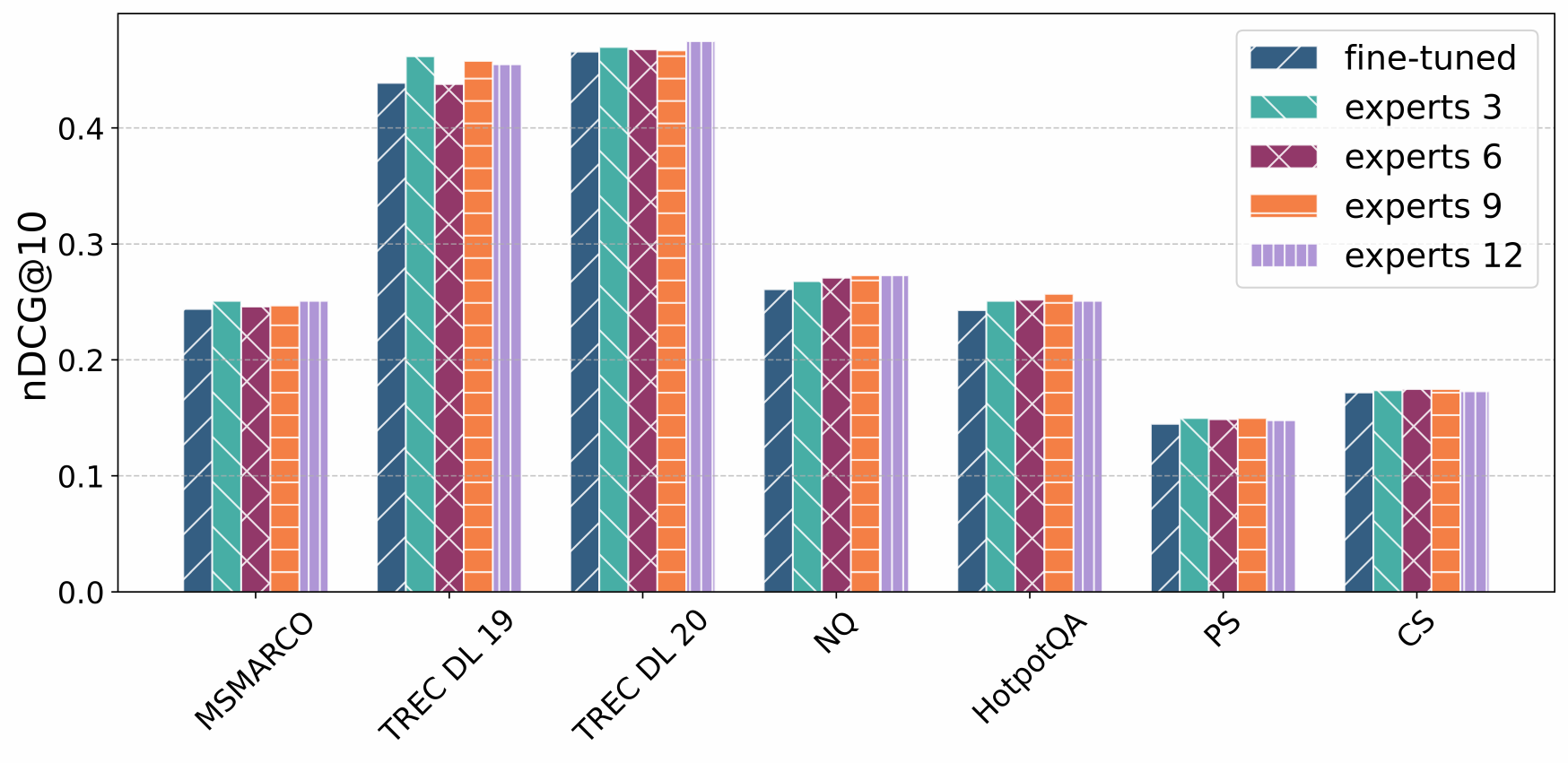}
    \end{subfigure}
    \hfill
    \begin{subfigure}[b]{0.47\textwidth}
        \includegraphics[width=\textwidth, height=5cm]{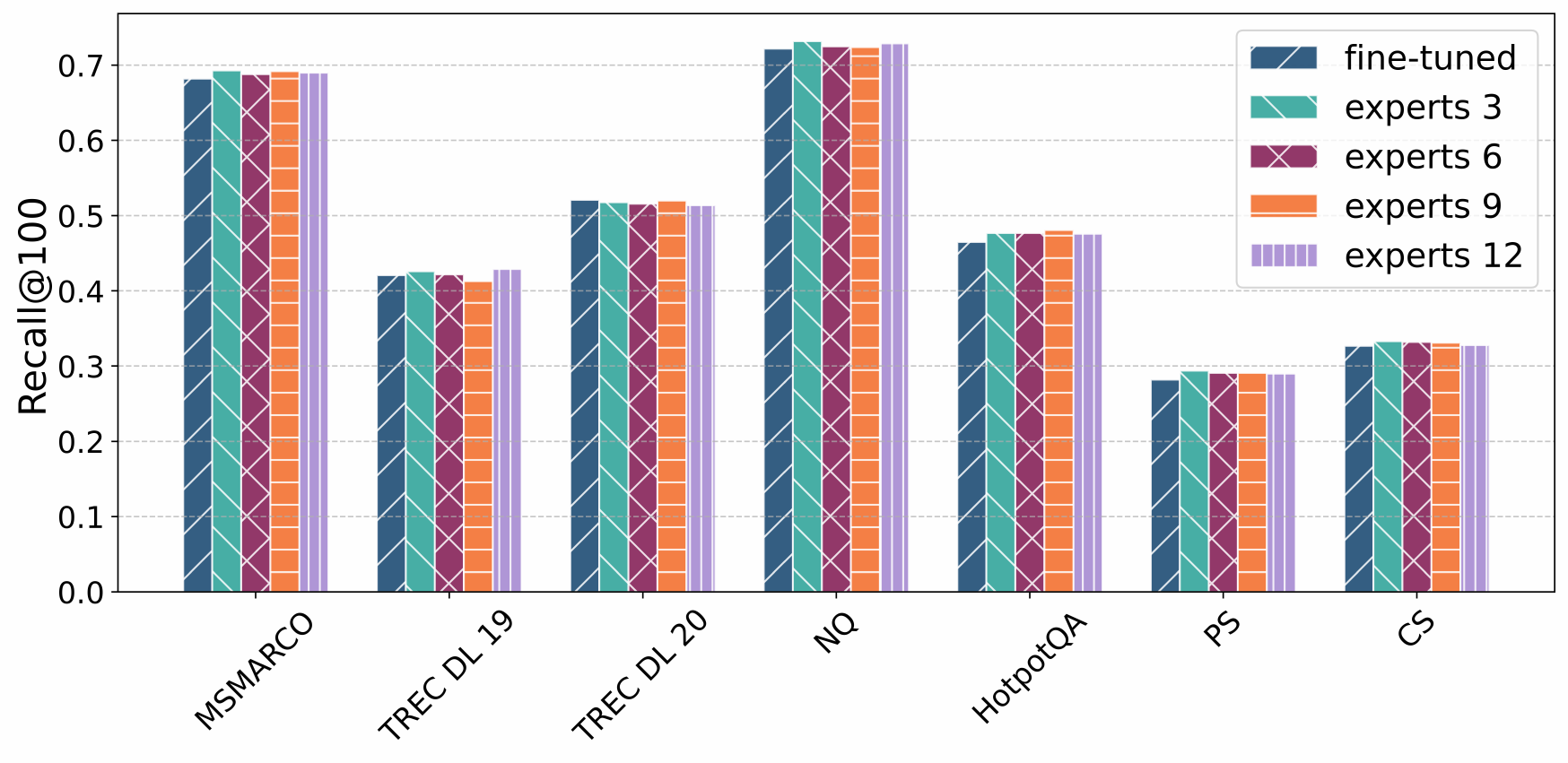}
    \end{subfigure}
    \caption{nDCG@10 and Recall@100 scores of \textsc{SB-MoE}$_{\text{ALL}}$ on TinyBERT with 3, 6, 9, and 12 experts.}
    \label{fig:employed}
\end{figure}
\label{sec:ablation}
As our approach appears to significantly benefit lightweight models, we further investigate its sensitivity to the number of employed and activated experts by conducting additional experiments with TinyBERT.
We evaluate MoE configurations with 3, 6, 9, and 12 experts across all datasets.

Figure \ref{fig:employed} shows the nDCG@10 and Recall@100 scores of \textsc{SB-MoE}$_{\text{ALL}}$\footnote{For these experiments, we use \textsc{SB-MoE}$_{\text{ALL}}$ as it outperforms \textsc{SB-MoE}$_{\text{TOP-1}}$ when integrated with TinyBERT on most of the evaluated benchmarks (for more details, refer to Tables~\ref{tab:results_v2} and \ref{tab:results_zero} analyzed in Section \ref{sec:results}).} compared to the performance of the fine-tuned version of TinyBERT on all seven evaluation benchmarks.
The results reveal that the retrieval performance varies considerably with changes in the number of employed experts.
For instance, on NQ and for nDCG@10, the performance gains range from 3.29\% to 5.76\% with different numbers of employed experts. 
Moreover, variations in the number of experts can lead to the maximization of different performance measures, as observed in the case of TREC DL 19, where using 3 experts maximizes nDCG@10, while Recall@100 is maximized with 12 experts.
In general, while at least one MoE configuration outperforms the fine-tuned baseline, no specific number of experts consistently yields optimal performance across all metrics and datasets.
Additionally, Figure \ref{fig:employed} illustrates that increasing the number of experts does not guarantee a performance improvement.
For example, in many cases, Recall@100 seems to be optimal with just 3 experts (e.g., MSMARCO, NQ, PS, and CS).
These observations indicate that calibrating this hyperparameter (i.e., the number of employed experts) can be effective in identifying the model's optimal performance for the downstream IR tasks.
\textsc{SB-MoE}, adding a single MoE block and having only a few additional parameters, allows for an efficient hyperparameter optimization.

To further assess performance fluctuations resulting from variations in the number of employed experts, we conduct a set of experiments to examine the number of \textit{activated} experts within each collection.
Table \ref{tab:results_active} shows the number of activated experts over the number of total experts employed for all five collections\footnote{The TREC DL 19 \& 20 benchmarks share a corpus with MSMARCO, hence are excluded from this set of experiments.}.
We consider activated those experts that were assigned the highest score by the gating function for at least 100k documents of the given corpus (over $1\%$ of the documents of the largest collection), as corpus sizes range from 2.8M to 8.8M documents.
During training, all MoE configurations (3 to 12 employed experts) used top-1 noisy gating, which enforces the model to explore all experts.
However, during inference, the obtained results show that the number of activated experts differs considerably across MoE configurations and collections.
In general, the number of activated experts ranges from 2 to 7, with a notable 100\% activation for the MoE configuration that employs 3 experts, where every expert is activated across all collections.
Moreover, we observe that increasing the number of employed experts does not necessarily increase the number of activated experts.
For example, in HotpotQA and for 9 employed experts, we observe the activation of 7 of them (77.78\%), but when we increase the number of employed experts to 12, the number of activated experts drops to 5 (41.67\%).
These findings highlight that the number of experts is a hyperparameter to be tuned for the downstream IR task, and, with no more than 7 experts ever activated per collection, performance gains can be achieved without significantly increasing computational overhead, as only a subset of experts is ultimately utilized.

\begin{table}[t]
\caption{The number of \textit{activated} experts during inference over the number of employed experts in \textsc{SB-MoE} on TinyBERT for all five collections. All four MoE configurations are trained with noisy Top-1 gating to ensure that all experts are explored during training. The \textit{"Threshold (\%)"} column indicates the minimum percentage of documents for each collection that an expert must process to be considered activated.}
\label{tab:results_active}
\centering
\resizebox{0.7\linewidth}{!}{%
\begin{tabular}{c|c|c|l}
\textbf{Corpus Size} & \textbf{Threshold (\%)} & \textbf{\#Experts} & \textbf{Activated} \\ \hline
\multirow{2}{*}{MSMARCO} & \multirow{4}{*}{$>1.14\%$} & 3   & 3 (100\%)  \\
                         & & 6   & 3 (50\%)        \\
\multirow{2}{*}{8.8M}    & & 9   & 6 (66.67\%)     \\
                         & & 12  & 5 (41.67\%)     \\ \hline
\multirow{2}{*}{NQ}      & \multirow{4}{*}{$>3.85\%$} & 3   & 3 (100\%)       \\
                         & & 6   & 2 (33.33\%)     \\
\multirow{2}{*}{2.6M}    & & 9   & 5 (55.56\%)     \\
                         & & 12  & 4 (33.33\%)     \\ \hline
\multirow{2}{*}{HotpotQA}& \multirow{4}{*}{$>1.92\%$} & 3   & 3 (100\%)       \\
                         & & 6   & 4 (66.67\%)     \\
\multirow{2}{*}{5.2M}    & & 9   & 7 (77.78\%)     \\
                         & & 12  & 5 (41.67\%)     \\ \hline
\multirow{2}{*}{PS}      & \multirow{4}{*}{$>2.08\%$} & 3   & 3 (100\%)       \\
                         & & 6   & 5 (83.33\%)     \\
\multirow{2}{*}{4.8M}    & & 9   & 5 (55.56\%)     \\
                         & & 12  & 5 (41.67\%)     \\ \hline
\multirow{2}{*}{CS}      & \multirow{4}{*}{$>2.08\%$} & 3   & 3 (100\%)       \\
                         & & 6   & 4 (66.67\%)     \\
 \multirow{2}{*}{4.8M}   & & 9   & 4 (44.44\%)     \\
                         & & 12  & 7 (58.33\%)     \\
\end{tabular}
}
\end{table}
\begin{figure}[t]
    \centering
    \includegraphics[width=1\linewidth]{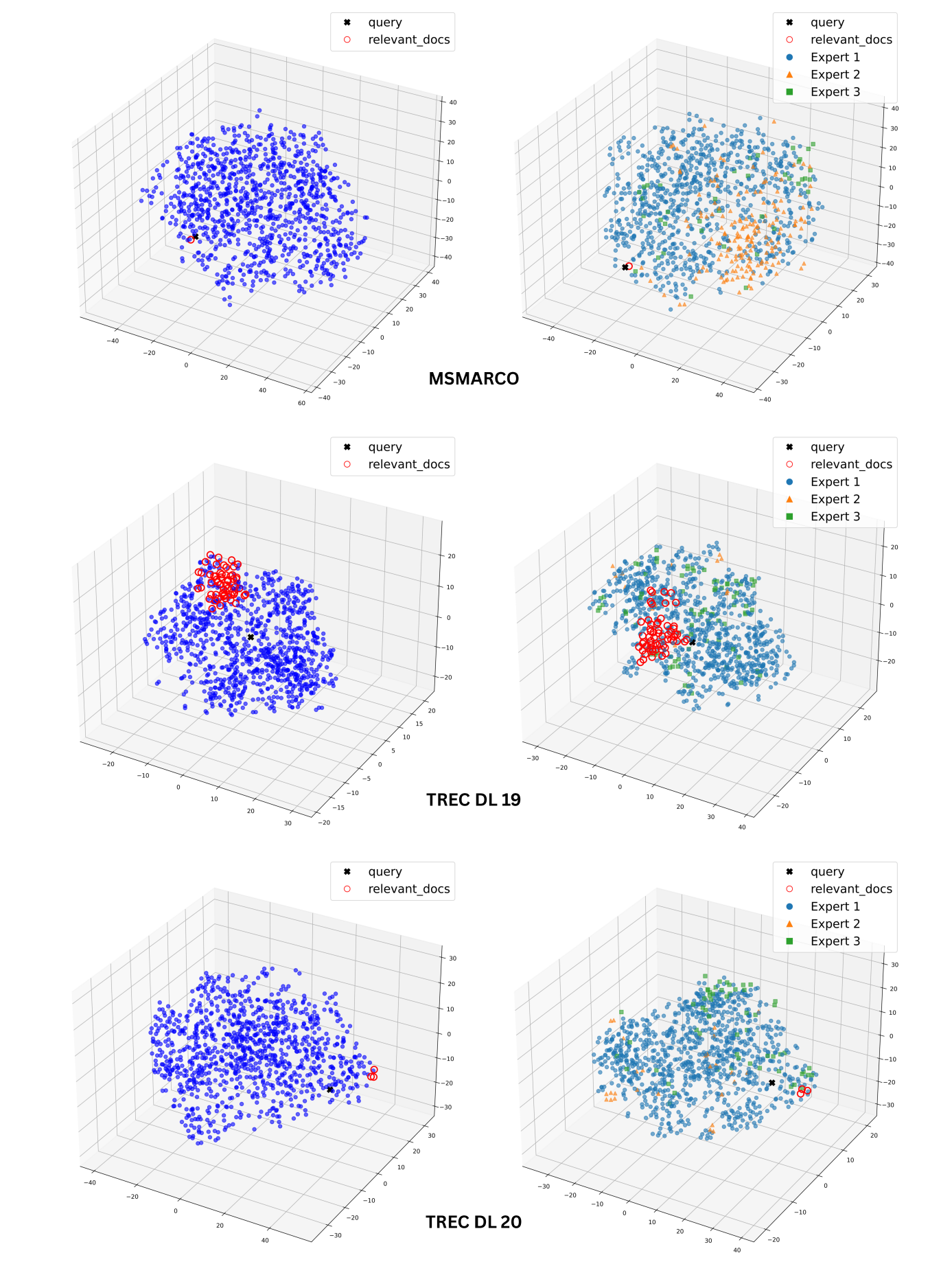}
    \caption{3D t-SNE visualizations of a query and its top 1000 retrieved documents, on the \textit{left} embedded by the \textit{original DRM} and on the \textit{right} by \textsc{\textit{SB-MoE}}. Color figure online.}
    \label{fig:tsne1}
\end{figure}
\begin{figure}[t]
    \centering
    \includegraphics[width=1\linewidth]{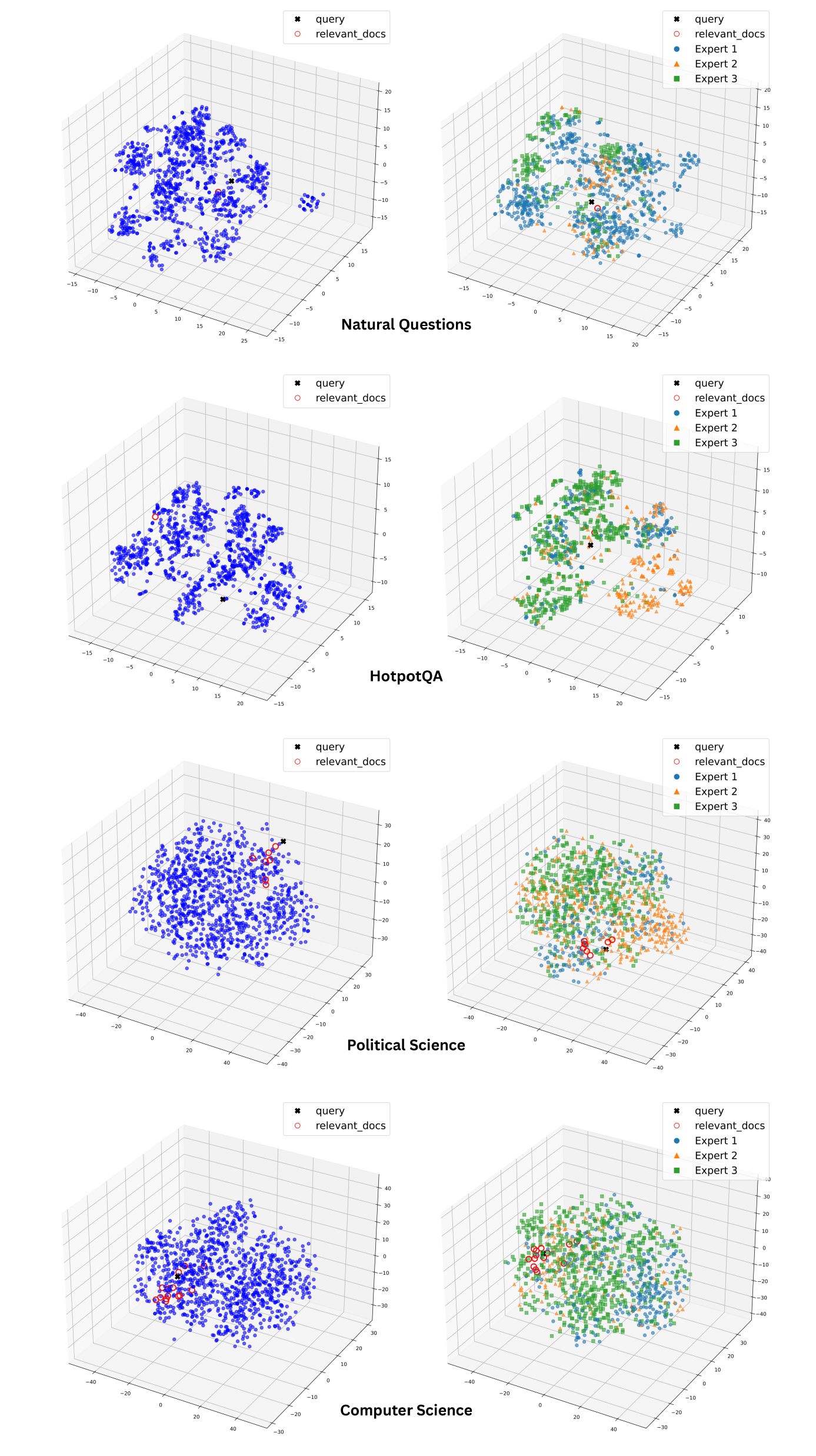}
    \caption{Continuation of Figure \ref{fig:tsne1}. 3D t-SNE visualizations of a query and its top 1000 retrieved documents, on the \textit{left} embedded by the \textit{original DRM} and on the \textit{right} by \textsc{\textit{SB-MoE}}.}
    \label{fig:tsne2}
\end{figure}

Figures \ref{fig:tsne1} \& \ref{fig:tsne2} illustrate examples\footnote{Due to space constraints, we opt to show one example per collection within the paper and present more in the provided code repository: \url{https://github.com/FaySokli/SB-MoE}.} of 3-dimensional t-SNE representations of queries and their top 1000 retrieved documents as these are embedded in the Dense Vector Space (DVS) by TinyBERT.
These illustrations show the formulation of the DVS before (Figures \ref{fig:tsne1} \& \ref{fig:tsne2} - \textit{left}) and after (Figures \ref{fig:tsne1} \& \ref{fig:tsne2} - \textit{right}) the addition of 3 experts across all seven benchmarks.
We choose to visualize the spaces after employing 3 experts, as in this MoE configuration, 100\% of the experts are activated for all evaluation collections.
Our findings indicate that:
\begin{enumerate}
    \item the given query and its relevant document(s) shift position drastically within the DVS after the application of the MoE block to the underlying DRM;
    \item documents processed by the same expert tend to form clusters within the DVS, suggesting that the experts are able to learn useful textual characteristics and inject them into the generated embeddings; and
    \item \textsc{SB-MoE} produces enriched query and document representations that align better within the DVS for similarity estimation, since there are noticeable improvements in the retrieval effectiveness of the underlying DRM with the addition of 3 experts (see Figure \ref{fig:employed}).
\end{enumerate}

\section{Conclusions}\label{sec:conclude}
In this work, we conduct an experimental investigation into the performance of an enhanced Dense Retrieval Model (DRM) architecture leveraging Mixture-of-Experts (MoE) to explore their potential in dense retrieval.
Specifically, we integrate a single MoE block (\textsc{SB-MoE}) into four different DRMs, jointly training it with the gating function and the underlying DRM in an unsupervised manner to enrich both query and document representations and automatically optimize them for the downstream dense retrieval tasks.
Results show that \textsc{SB-MoE} significantly enhances retrieval performance when integrated with DRMs with lightweight base models, consistently improving nDCG@10 and Recall@100 across all benchmarks.
However, only marginal gains were noted for larger DRMs, suggesting that \textsc{SB-MoE} may be less impactful in DRMs with high parameter counts.
Our findings also highlight the critical role of the learned gating function in expert selection within MoE approaches; the addition of experts alone is insufficient, and retrieval performance gains in DRMs rely on effective, optimized expert routing.
Additional empirical evaluations show that \textsc{SB-MoE} can generalize well across domains and tasks while maintaining low computational overhead, making it well-suited for low-resource environments.
Further analyses on the number of employed and activated experts reveal that these hyperparameters have a significant impact on the proposed model's performance.
Since only a subset of experts is effectively utilized, performance gains can be achieved with minimal impact on efficiency.

\section*{Acknowledgments}
This work has received funding from the European Union’s Horizon Europe research and innovation programme under the Marie Skłodowska-Curie grant agreement No 101073307.

\bibliographystyle{IEEEtran}
\bibliography{refs}

\end{document}